\documentclass[aps,pra,reprint]{revtex4-1}
\usepackage{blindtext}
\usepackage{amssymb}
\usepackage{amsmath}
\usepackage{tikz}
\usepackage{graphicx}
\usepackage{siunitx}

\newtheorem{prop}{Proposition}\def\PRO{\begin{prop}}\def\ORP{\end{prop}}
\newtheorem{coro}{Corollary}\def\COR{\begin{coro}}\def\ROC{\end{coro}}
\newtheorem{theo}{Theorem}\def\TH{\begin{theo}}\def\HT{\end{theo}}
\def\TH{\begin{theo}}\def\HT{\end{theo}}
\newtheorem{defi}[prop]{Definition}\def\DE{\begin{defi}}\def\ED{\end{defi}}

\newtheorem{lemme}[prop]{Lemma}\def\LE{\begin{lemme}}\def\EL{\end{lemme}}
\usepackage{color}
\usepackage{graphicx,float}
\usepackage{epsfig}
\usepackage{amsmath,amssymb}
\usepackage{bm}
\usepackage[latin1]{inputenc}
\usepackage{hyphenat}
\usepackage[bookmarks=false]{hyperref}
\usepackage{color}
\usepackage[capitalise]{cleveref}
\usepackage{epstopdf}
\usepackage{braket}
\usepackage{verbatim}
\usepackage{comment}
\usepackage{multirow}
\usepackage{subcaption}
\usepackage{float}

\def\ket#1{\left| #1 \right\rangle}

\newcommand{\beq}{\begin{equation}}
\newcommand{\eeq}{\end{equation}}

\definecolor{pink}{RGB}{255,0,255}

\definecolor{ss_color}{rgb}{0,0,1}

\definecolor{darkorange}{RGB}{255,120,0} 

\definecolor{red}{rgb}{1,0,0}

\begin{document}

\title{Insecurity of detector-device-independent quantum key distribution}
\author{Shihan~Sajeed$^{1,2}$}
\email{shihan.sajeed@gmail.com}
\author{Anqi~Huang$^{1,2}$}
\author{Shihai~Sun$^3$}
\author{Feihu Xu$^4$}
\author{Vadim~Makarov$^{1,2,5}$}
\author{Marcos Curty$^6$}
\affiliation{$^1$Institute for Quantum Computing, University of Waterloo, Waterloo, ON, N2L~3G1 Canada\\
$^2$Department of Electrical and Computer Engineering, University of Waterloo, Waterloo, ON, N2L~3G1 Canada\\
$^3$College of Science, National University of Defense Technology, Changsha 410073, China\\
$^4$Research Laboratory of Electronics, Massachusetts Institute of Technology, 77 Massachusetts Avenue, Cambridge, MA 02139, USA\\
$^5$Department of Physics and Astronomy, University of Waterloo, Waterloo, ON, N2L~3G1 Canada\\
$^6$Escuela de Ingenier\'ia de Telecomunicaci\'on, Department of Signal Theory and Communications, University of Vigo, Vigo E-36310, Spain
}

\date{\today}

\begin{abstract}
Detector-device-independent quantum key distribution (ddiQKD) held the promise of being robust to detector side-channels, a major security loophole in QKD implementations. In contrast to what has been claimed, however, we demonstrate that the security of ddiQKD is not based on post-selected entanglement, and we introduce various eavesdropping strategies that show that ddiQKD is in fact insecure against detector side-channel attacks as well as against other attacks that exploit device's imperfections of the receiver. Our attacks are valid even when the QKD apparatuses are built by the legitimate users of the system themselves, and thus free of malicious modifications, which is a key assumption in ddiQKD.
\end{abstract}

\maketitle

\noindent {\it Introduction.}---Quantum key distribution (QKD), a technique to distribute a secret random bit string between two separated parties (Alice and Bob), needs to close the gap between theory and practice~\cite{lo2014}. In theory, QKD provides information-theoretic security. In practice, however, it does not because QKD implementation devices do not typically conform to the theoretical models considered in the security proofs. As a result, any unaccounted device imperfection might constitute a side-channel which could be used by an eavesdropper (Eve) to learn the secret key without being detected~\cite{vakhitov2001,makarov2006,qi2007,lamas-linares2007,lydersen2010a,xu2010,gerhardt2011,weier2011,jouguet2013,sajeed2015,sajeed2015a}.

To bridge this gap, various approaches have been proposed recently~\cite{mayers1998,acin2007,vazirani2014,miller2014,lo2012}, with measurement-device-independent QKD (mdiQKD)~\cite{lo2012} probably being the most promising one in terms of feasibility and performance. Its security is based on post-selected entanglement, and it can remove all detector side-channels from QKD implementations, which is arguably their major security loophole~\cite{makarov2006,qi2007,lamas-linares2007,lydersen2010a,xu2010,gerhardt2011,weier2011,jouguet2013,sajeed2015a}. Also, its practicality has been already confirmed both in laboratories and via field trials~\cite{rubenok2013,silva2013,liu2013,tang2014,comandar2016,tang2016,yin2016}. A drawback of mdiQKD is, however, that it requires high-visibility two-photon interference between independent sources, which makes its implementation more demanding than that of conventional QKD schemes. In addition, current finite-key security bounds against general attacks~\cite{curty2014} require larger post-processing data block sizes than those of standard QKD.

To overcome these limitations, a novel approach, so-called detector-device-independent QKD (ddiQKD), has been introduced recently~\cite{gonzalez2015,lim2014,cao2014,liang2015}. It avoids the problem of interfering photons from independent light sources by using the concept of a single-photon Bell state measurement (BSM)~\cite{kim2003}. As a result, its finite-key security bounds and classical post-processing data block sizes are expected to be similar to those of prepare-and-measure QKD schemes~\cite{lim2013}. Despite this presumed promising performance, however, the robustness of ddiQKD against detector side-channel attacks has not been rigorously proven yet, and only partial security proofs have been introduced~\cite{gonzalez2015,lim2014}.

Here, we show that, in contrast to what has been claimed~\cite{gonzalez2015,lim2014,cao2014,liang2015}, the security of ddiQKD {\it cannot} rely on the same principles as mdiQKD ({\it i.e.,}\ post-selected entanglement). More importantly, we demonstrate that ddiQKD is in fact vulnerable to detector side-channel attacks and to other attacks that exploit imperfections of the receiver's devices. These attacks are valid even when Alice's and Bob's state preparation processes are fully characterised and trusted, an essential assumption in ddiQKD. Moreover, they do not require that Eve substitutes Bob's detectors with a measurement apparatus prepared by herself to leak key information to the channel~\cite{qi2015}. That is, our attacks apply as well to the scenario where Alice and Bob build the QKD devices themselves.

\noindent {\it mdiQKD \& ddiQKD.}---Let us start by reviewing the basic principles behind mdiQKD and ddiQKD. To simplify the discussion, we shall assume that Alice and Bob have at their disposal perfect single-photon sources. Note, however, that both schemes can operate as well, for instance, with phase-randomised weak coherent pulses in combination with decoy states~\cite{hwang2003,lo2005,wang2005a}.

\begin{figure}[t]
  \includegraphics[width=0.82\columnwidth]{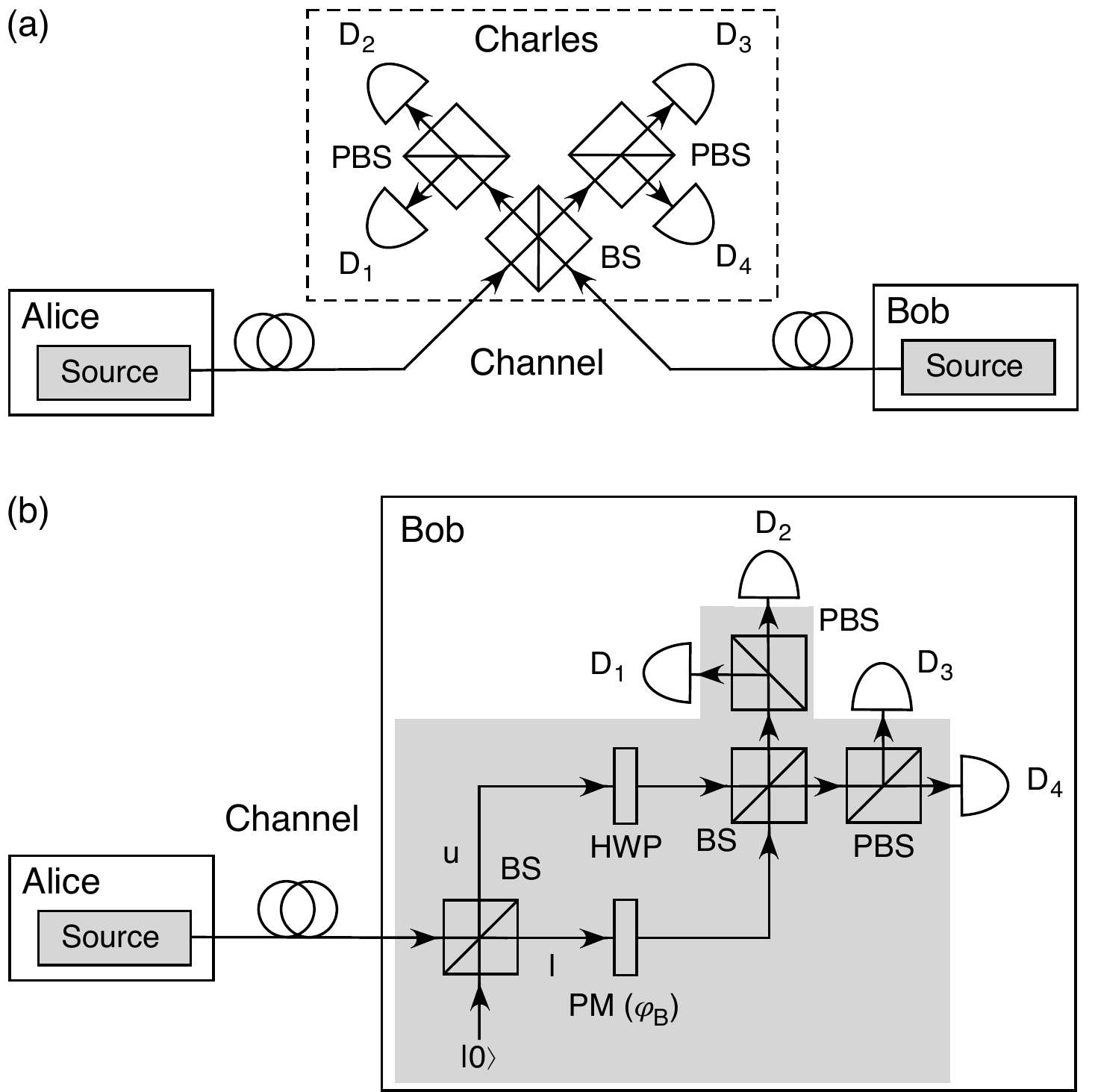}
\caption{Possible implementations of partially-device-independent QKD with linear optics. (a)~mdiQKD~\cite{lo2012}. PBS, polarising beamsplitter; BS, 50\,:\,50 beamsplitter;
and D$_i$, with $i\in\{1,2,3,4\}$, Charles' single-photon detectors. (b)~ddiQKD~\cite{lim2014}. HWP, half-wave plate; and PM, phase modulator. One single click in the detector D$_1$, D$_2$, D$_3$, or D$_4$ corresponds to a projection into the Bell state $\ket{\Psi^+}$, $\ket{\Phi^+}$, $\ket{\Psi^-}$, or $\ket{\Phi^-}$ respectively (see main text for further details). In both schemes, the grey areas denote devices that need to be characterised and trusted. Also, Alice's and Bob's laboratories need to be protected from any information leakage to the outside.}
\label{mdiQKD_vs_ddiQKD}
\end{figure}

An example of a possible implementation of mdiQKD is illustrated in Fig.~\ref{mdiQKD_vs_ddiQKD}(a)~\cite{lo2012}. Both Alice and Bob generate BB84 states~\cite{bennett1984} and send them to an untrusted relay Charles. If Charles is honest, he performs a two-photon BSM that projects the incoming signals into a Bell state. In any case, Charles has to declare which of his measurements are successful together with the Bell states obtained. Alice and Bob then extract a secret key from those successful events where they used the same basis. Importantly, if Charles is honest, his BSM measurement post-selects entanglement between Alice and Bob, and, therefore, he is not able to learn any information about their bit values. To test whether or not Charles is honest, Alice and Bob can simply compare a randomly chosen subset of their data to see if it satisfies the expected correlations associated to the Bell states announced. That is, mdiQKD can be seen as a time-reversed Einstein-Podolsky-Rosen QKD protocol~\cite{biham1996}. Therefore, its security can be proven without any assumption on the behaviour of Charles' measurement unit.

ddiQKD~\cite{gonzalez2015,lim2014,cao2014,liang2015} aims to follow the same spirit of mdiQKD. The key idea is to replace the two-photon BSM with a two-qubit single-photon BSM~\cite{kim2003}. This requires that Alice and Bob use two different degrees of freedom of the single-photons to encode their bit information. In so doing, one avoids the need for interfering photons from independent light sources. An example of a possible implementation is illustrated in~Fig.~\ref{mdiQKD_vs_ddiQKD}(b)~\cite{lim2014} (see also~\cite{gonzalez2015,cao2014,liang2015} for similar proposals). Here, Alice sends Bob BB84 polarisation states: $(\ket{\rm{H}}+e^{i\theta_{\rm A}}\ket{\rm{V}})/\sqrt{2}$, where $\ket{\rm{H}}$ ($\ket{\rm{V}}$) denotes the Fock state of a single-photon prepared in horizontal (vertical) polarisation, and the phase $\theta_{\rm A}\in\{0,\pi/2,\pi, 3\pi/2\}$. Bob then encodes his bit information by using the spatial degree of freedom of the incoming photons. This is done with a 50\,:\,50 beamsplitter (BS) together with a phase modulator (PM) that applies a random phase $\varphi_{\rm B}\in\{0,\pi/2,\pi, 3\pi/2\}$ to each input signal. Finally, Bob performs a BSM that projects each input photon into a Bell state: $\ket{\Phi^\pm}=(\ket{\rm{H}}\ket{\rm{u}}\pm\ket{\rm{V}}\ket{\rm{l}})/\sqrt{2}$ and $\ket{\Psi^\pm}=(\ket{\rm{H}}\ket{\rm{l}}\pm\ket{\rm{V}}\ket{\rm{u}})/\sqrt{2}$, where $\ket{\rm{u}}$ ($\ket{\rm{l}}$) represents the state of a photon that goes through the upper (lower) arm of the interferometer (see~Fig.~\ref{mdiQKD_vs_ddiQKD}(b)). A photon detection event (``click'') in only one detector ${\rm D}_i$ corresponds to a projection on a particular Bell state.

Both mdiQKD and ddiQKD require that Alice's and Bob's state preparation processes are characterised and trusted. This is indicated by the grey areas shown in Fig.~\ref{mdiQKD_vs_ddiQKD}. In ddiQKD, the elements inside Bob's grey area can be regarded as his trusted transmitter (when compared to mdiQKD). Among the trusted components there are elements which belong to the BSM, but, importantly, the detectors ${\rm D}_i$ do not need to be trusted.

\noindent {\it The security of ddiQKD is not based on post-selected entanglement.}---At a first sight, it seems that the security of ddiQKD follows directly from that of mdiQKD, given, of course, that the assumptions on Alice's and Bob's state preparation processes are fulfilled~\cite{gonzalez2015,lim2014,cao2014,liang2015}. That is, it relies on the fact that the BSM post-selects entanglement between Alice and Bob. A first indication that confronts this idea was given recently in~\cite{qi2015}. There, it was shown that, in contrast to mdiQKD,  ddiQKD is actually insecure if Eve is able to replace Bob's detectors with a measurement apparatus that leaks information to the channel~\cite{qi2015}. Although this result is important from a conceptual point of view, it violates one of the security assumptions of ddiQKD: Bob's detectors have to be built by a trusted party (but do not need to be characterised) to avoid that they intentionally leak key information to the outside~\cite{gonzalez2015}. Below we show that even in this scenario, the security of ddiQKD cannot be based on post-selected entanglement alone, unlike mdiQKD.

For this, we will consider a slightly simplified version of the ddiQKD scheme illustrated in~Fig.~\ref{mdiQKD_vs_ddiQKD}(b). In particular, we will assume that Bob's receiver has only one active detector, say for instance the detector $\rm{D}_1$, while the other detectors are disabled. That is, now Bob's BSM projects the incoming photons only into the Bell state $\ket{\Psi^+}$. If the security of ddiQKD is based on post-selected entanglement, this modification should not affect its security (only its secret key rate is reduced by a factor of four), as a projection into a single Bell state should be sufficient to guarantee security~\cite{lo2012}. Next we show that a blinding attack~\cite{lydersen2010a,gerhardt2011} renders ddiQKD insecure in this situation.

In particular, suppose that Eve shines bright light onto Bob's detector $\rm{D}_1$ to make it enter linear-mode operation~\cite{lydersen2010a,gerhardt2011}. In this mode the detector is no longer sensitive to single-photon pulses, but it can only detect strong light. In order to simplify the discussion, we shall assume that $\rm{D}_1$ always produces a click (no click) outcome when the mean photon number $\mu$ of the incoming pulses satisfies $\mu\geq\mu_{\rm th}$ ($\mu<\mu_{\rm th}$) for a certain threshold value $\mu_{\rm th}$. Once $\rm{D}_1$ is blinded, Eve performs an intercept-resend attack on every signal sent by Alice. That is, she measures Alice's signals in one of the two BB84 bases (which Eve selects at random for each pulse), and she prepares a new signal, depending on the result obtained, that is sent to Bob. Intercept-resend attacks correspond to entanglement-breaking channels and, therefore, they cannot lead to a secure key~\cite{curty2004a}. Suppose, for instance, that the signals that Eve sends to Bob are coherent states of the form $\ket{\sqrt{2\mu}}$ with creation operator $a^\dagger=(a_{\rm H}^\dagger+e^{i\phi_{\rm E}}a_{\rm V}^\dagger)/\sqrt{2}$. Here, $a_{\rm H}^\dagger$ ($a_{\rm V}^\dagger$) denotes the creation operator for horizontally (vertically) polarised photons, and the phase $\phi_{\rm E}\in\{0,\pi/2,\pi, 3\pi/2\}$ depends on Eve's measurement result. More precisely, for each measured signal, Eve sends Bob a coherent state prepared in the BB84 polarisation state identified by her measurement. Then, it can be shown that the state at the input ports of Bob's detectors $\rm{D}_i$ is a coherent state of the form (see Appendix~\ref{calc_main_eq} for details)
\begin{eqnarray}\label{main_eq}
\ket{\psi}&=&\ket{\frac{\sqrt{\mu}}{2}\big(e^{i\phi_{\rm E}}+e^{i\varphi_{\rm B}}\big)}_{\rm{D}_1}\otimes
\ket{\frac{\sqrt{\mu}}{2}\big(1+e^{i(\phi_{\rm E}+\varphi_{\rm B})}\big)}_{\rm{D}_2} \nonumber \\
&\otimes&\ket{\frac{\sqrt{\mu}}{2}\big(e^{i\phi_{\rm E}}-e^{i\varphi_{\rm B}}\big)}_{\rm{D}_3}
\otimes\ket{\frac{\sqrt{\mu}}{2}\big(1-e^{i(\phi_{\rm E}+\varphi_{\rm B})}\big)}_{\rm{D}_4}. \nonumber \\
\end{eqnarray}

\begin{table}
  \caption{Mean photon number of the input light to Bob's detectors as a function of the phases $\phi_{\rm E}$ and $\varphi_{\rm B}$.}
  \label{tbl:main}
  \begin{subtable}[t]{0.5\linewidth}
     \caption{$\phi_{\rm E}=0$}
     \label{tbl:a}
    \begin{tabular}{c|cccc}
       $\varphi_{\rm B}$ & D$_1$ & D$_2$ &  D$_3$ & D$_4$ \\
       \hline
       0 & $\mu$ & $\mu$  & 0 & 0 \\
       $\frac{\pi}{2}$ & $\frac{\mu}{2}$ & $\frac{\mu}{2}$  & $\frac{\mu}{2}$ & $\frac{\mu}{2}$  \\
       $\pi$ & 0 & 0  & $\mu$ & $\mu$ \\
       $\frac{3\pi}{2}$ & $\frac{\mu}{2}$ & $\frac{\mu}{2}$  & $\frac{\mu}{2}$ & $\frac{\mu}{2}$ \\
     \end{tabular}
     \vspace{1.4\baselineskip}

    \caption{$\phi_{\rm E}=\frac{\pi}{2}$}
     \label{tbl:b}
     \begin{tabular}{c|cccc}
       $\varphi_{\rm B}$ & D$_1$ & D$_2$ &  D$_3$ & D$_4$ \\
       \hline
       0 & $\frac{\mu}{2}$ & $\frac{\mu}{2}$  & $\frac{\mu}{2}$ & $\frac{\mu}{2}$ \\
       $\frac{\pi}{2}$ & $\mu$ & 0  & 0 & $\mu$  \\
       $\pi$ & $\frac{\mu}{2}$ & $\frac{\mu}{2}$  & $\frac{\mu}{2}$ & $\frac{\mu}{2}$ \\
       $\frac{3\pi}{2}$ & 0 & $\mu$  & $\mu$ & $0$ \\ 
    \end{tabular}
    \end{subtable}\hfill
  \begin{subtable}[t]{0.5\linewidth}

    \caption{$\phi_{\rm E}=\pi$}
    \label{tbl:c}
   \begin{tabular}{c|cccc}
  $\varphi_{\rm B}$ & D$_1$ & D$_2$ &  D$_3$ & D$_4$ \\
  \hline
   0 & 0 & 0  & $\mu$ & $\mu$ \\
  $\frac{\pi}{2}$ & $\frac{\mu}{2}$ & $\frac{\mu}{2}$  & $\frac{\mu}{2}$ & $\frac{\mu}{2}$  \\
  $\pi$ & $\mu$ & $\mu$  & 0 & 0 \\
  $\frac{3\pi}{2}$ & $\frac{\mu}{2}$ & $\frac{\mu}{2}$  & $\frac{\mu}{2}$ & $\frac{\mu}{2}$ \\ 
    \end{tabular}
     \vspace{1.4\baselineskip}
 
    \caption{$\phi_{\rm E}=\frac{3\pi}{2}$}
    \label{tbl:d}
     \begin{tabular}{c|cccc}
  $\varphi_{\rm B}$ & D$_1$ & D$_2$ &  D$_3$ & D$_4$ \\
  \hline
   0 & $\frac{\mu}{2}$ & $\frac{\mu}{2}$  & $\frac{\mu}{2}$ & $\frac{\mu}{2}$ \\
  $\frac{\pi}{2}$ & 0 & $\mu$  & $\mu$ & $0$  \\
  $\pi$ & $\frac{\mu}{2}$ & $\frac{\mu}{2}$  & $\frac{\mu}{2}$ & $\frac{\mu}{2}$ \\
  $\frac{3\pi}{2}$ & $\mu$ & 0  & 0 & $\mu$ \\ 
    \end{tabular}
  \end{subtable}
\end{table}

This situation is illustrated in Table~\ref{tbl:main}, where we show the mean photon number of the incoming light to Bob's detectors for all combinations of $\phi_{\rm E}$ and $\varphi_{\rm B}$. Most importantly, from this table we can see that if $\rm{D}_1$ is the only active detector and Eve selects $\mu$ such that $\mu/2<\mu_{\rm th}<\mu$, then Bob can only obtain a click when he uses the same measurement basis as Eve ({\it i.e.,}\ when $\varphi_{\rm B}, \phi_{\rm E}\in\{0,\pi\}$ or $\varphi_{\rm B},\phi_{\rm E}\in\{\pi/2,3\pi/2\}$). That is, this attack does not introduce any error. Moreover, we have that Bob and Eve select the same basis with at least $1/2$ probability. This means that the ddiQKD scheme illustrated in~Fig.~\ref{mdiQKD_vs_ddiQKD}(b) (with only one active detector) is actually insecure against the detector blinding attack for a total system loss beyond only $3$~dB, just like standard QKD schemes. This confirms that the security of ddiQKD cannot be based on post-selected entanglement. The same conclusion applies as well to the ddiQKD schemes introduced in Refs.~\cite{gonzalez2015},~\cite{cao2014}, and~\cite{liang2015}.

\noindent {\it Insecurity of ddiQKD against detector side-channel attacks.}---If Bob uses four active detectors, the detector blinding attack has one main drawback: it produces double-clicks~\cite{qi2015}. From Table~\ref{tbl:main} one can already see that whenever Bob uses the same measurement basis as Eve there is always two detectors that click. For instance, when $\varphi_{\rm B}=\phi_{\rm E}=0$ the detectors ${\rm D}_1$ and ${\rm D}_2$ always click, and similar for the other cases. This means that Alice and Bob could, in principle, try to monitor double-clicks to detect the presence of Eve. So, the question is whether or not four active detectors can make ddiQKD secure again. As we show below, the answer is ``no''. For this, we introduce two possible eavesdropping strategies that exploit practical imperfections of Bob's detectors to avoid double-clicks. See also Appendix~\ref{sca_linearopticsnetwork} for two alternative attacks that achieve the same goal by exploiting other imperfections of Bob's linear optics network.

\begin{figure}
  \includegraphics[width=0.85\columnwidth]{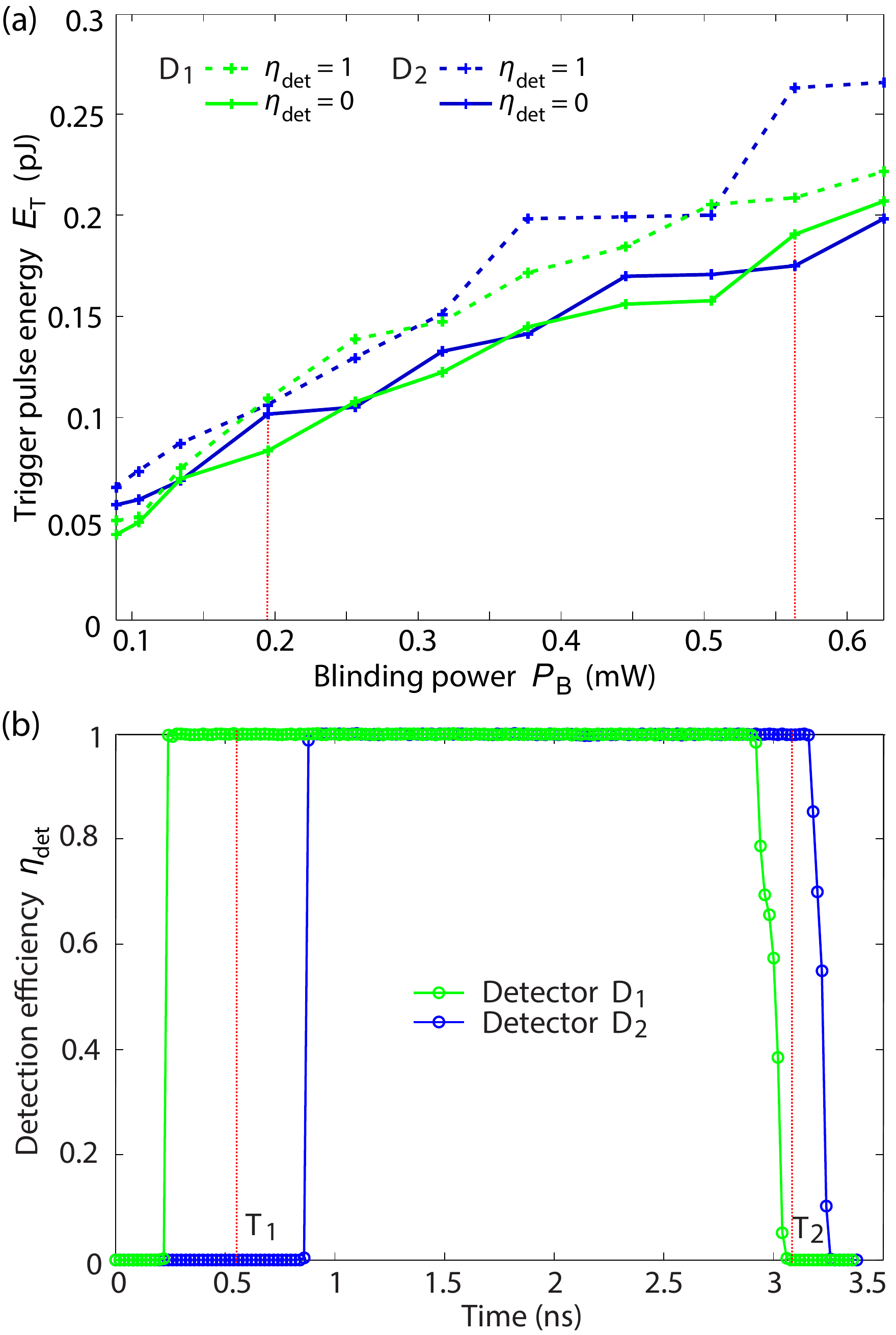}
  \caption{Detector click probability in bright-light blinded regime in commercial QKD system Clavis2. (a)~Click trigger thresholds versus blinding power $P_{\rm B}$ for two different single-photon detectors ${\rm D}_1$ and ${\rm D}_2$. Here, for a particular blinding power $P_{\rm B}$, each point in the solid (dashed) curves represents the maximum (minimum) value of trigger pulse energy $E_{\rm T}$ for which the detection efficiency $\eta_{\rm det}$ is $0$ $(1)$. The experimental data has been reprinted from Ref.~\cite{huang2016}. (b)~Measured detection efficiency mismatch in the time domain between two blinded single-photon detectors at $P_{\rm B} = \SI {0.32}{\milli\watt}$, $E_{\rm T} = \SI{0.24}{\pico\joule}$, and $\SI{0.7}{\nano\second}$ wide trigger pulse (see main text for further details).}
  \label{blinding}
\end{figure}

The first eavesdropping strategy uses the fact that single-photon detectors respond differently to the same blinding power $P_{\rm B}$. This has been recently analysed in Ref.~\cite{huang2016}. There, the authors compare the response of two single-photon detectors in a commercial QKD system Clavis2~\cite{idqclavis2specs} to varying blinding power. They first illuminate the detectors with continuous-wave bright light of power $P_{\rm B}$ to force them enter linear-mode operation. Then they record the maximum and minimum value of the trigger pulse energy $E_{\rm T}$ for which the click probabilities are $0$ and $1$ respectively. The results are shown in~Fig.~\ref{blinding}(a)~\cite{huang2016}. For a particular blinding power $P_{\rm B}$, each point in the solid (dashed) curves shown in the figure represents the maximum (minimum) value of trigger pulse energy $E_{\rm T}$ for which the detection efficiency $\eta_{\rm det}$ is $0$ $(1)$. The blue and green colours identify the two detectors. Next, we show how this effect could be used to avoid double-clicks.

For this, we return to the blinding attack described above against the ddiQKD implementation illustrated in Fig.~\ref{mdiQKD_vs_ddiQKD}(b). For simplicity, let us consider again the case where $\varphi_{\rm B}=\phi_{\rm E}=0$. In particular, suppose for instance that Eve wants to force a click only on detector say ${\rm D}_1$, and no click on detector ${\rm D}_2$. Then, in order to achieve this goal, she can simply choose a combination of $P_{\rm B}$ and $E_{\rm T}$ such that the detector ${\rm D}_1$ (${\rm D}_2$) has a non-zero (zero) click probability. If the behaviour of the detector ${\rm D}_1$ (${\rm D}_2$) corresponds to the green (blue) curves shown in~Fig.~\ref{blinding}(a), then the values $P_{\rm B} \approx 0.2$~mW and $E_{\rm T} \approx 0.1$~pJ constitute an example that satisfies this criterion. Similarly, if $P_{\rm B} \approx 0.56$~mW and $E_{\rm T} \approx  0.19$~pJ, then Eve could make the detector ${\rm D}_2$ (${\rm D}_1$) to have a non-zero (zero) click probability. Importantly, note that when Bob's basis matches that of Eve, only two out of the four detectors ${\rm D}_i$ might produce a click (see Table~\ref{tbl:main}). Hence, in these instances Eve only needs to avoid double-clicks between two detectors in order to remain undetected. A similar argument can be applied as well to any other value of $\varphi_{\rm B}$ and $\phi_{\rm E}$.

This attack demonstrates that if Bob's detectors are uncharacterised, as assumed in ddiQKD, this type of schemes are indeed insecure against detector side-channel attacks. That is, Eve could learn the whole secret key without producing any error nor a double-click.

A second eavesdropping strategy that also allows Eve to avoid double-clicks is based on a time-shift attack~\cite{makarov2006,qi2007} that exploits the detection efficiency mismatch between Bob's detectors. In this type of attack, Eve shifts the arrival time of each signal that she sends to Bob such that only one detector can produce a click each given time. Here, we have confirmed experimentally that this type of attack is also possible with blinded detectors. For this, we blinded two single-photon detectors from the commercial QKD system Clavis2~\cite{idqclavis2specs} and we measured their detection efficiency mismatch. The experimental results are shown in~Fig.~\ref{blinding}(b). We find, for instance, that whenever Bob receives a trigger pulse at the time instance $T_1$ ($T_2$), only the detector ${\rm D}_1$ (${\rm D}_2$) can produce a click because this instance is outside of the response region of the detector ${\rm D}_2$ (${\rm D}_1$). That is, by combining the time-shift attack with the blinding attack introduced in the previous section, Eve could again break the security of ddiQKD without introducing errors nor double-clicks.

\noindent {\it Conclusion.}---We have analysed the security of detector-device-independent QKD (ddiQKD), a novel scheme that promised to be robust against detector side-channel attacks. We have shown that its security is not based on post-selected entanglement, as originally claimed. Most importantly, we have presented various eavesdropping attacks that demonstrate that ddiQKD is actually vulnerable to detector side-channel attacks as well as to other side-channel attacks that exploit imperfections of Bob's receiver. These attacks are valid even when Alice's and Bob's state preparation processes are fully characterised and trusted, and Bob's detectors are built by a trusted party and cannot be replaced with a measurement device manufactured by Eve. 

\noindent {\it Acknowledgments.}---This work was supported by Industry Canada, CFI, NSERC (programs Discovery, PDF, CryptoWorks21), Ontario MRI, US Office of Naval Research, National Natural Science Foundation of China (grant No.\ 11304391), Spain MINECO, FEDER (grant No.\ TEC2014-54898-R), and Galician Regional Government (programs EM2014/033, AtlantTIC). The authors thank ID~Quantique for cooperation, technical assistance, and providing the QKD hardware.

\appendix

\section{Quantum state $\bm{\ket{\psi}}$ at the input ports of Bob's detectors}\label{calc_main_eq}

\begin{figure}
  \includegraphics[width=0.75\columnwidth]{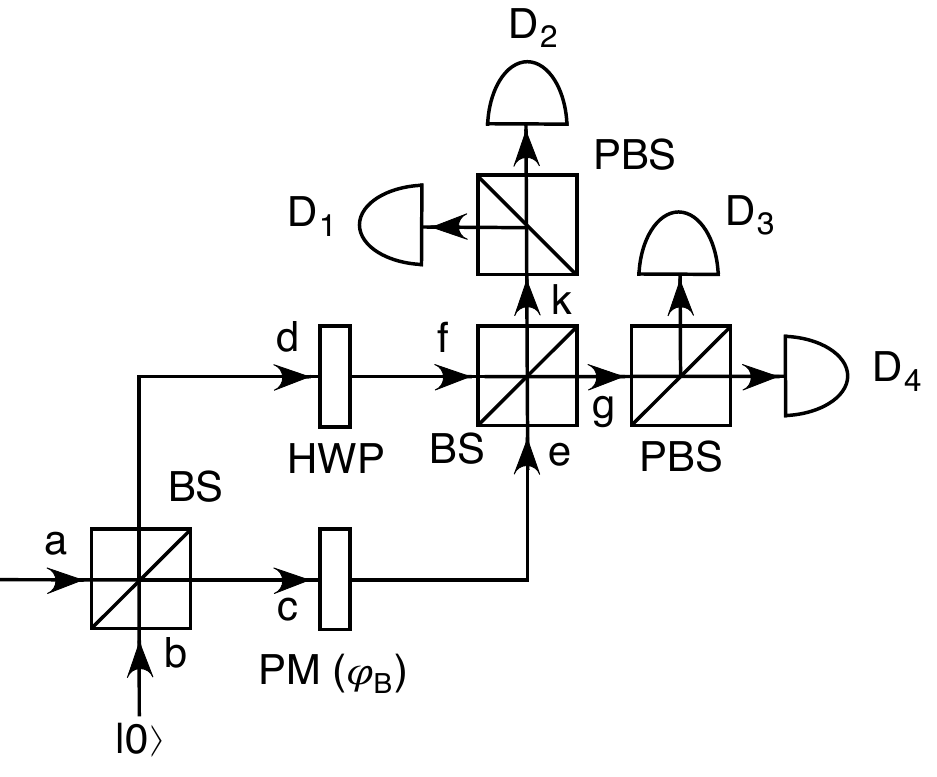}
\caption{Schematic representation of Bob's ddiQKD receiver. The different modes $a$, $b$, $c$, $d$, $e$, $f$, $g$, and $k$ correspond to those considered in the calculations of Appendix~\ref{calc_main_eq}; the receiver scheme is otherwise identical to Fig.~\ref{mdiQKD_vs_ddiQKD}(b)~\cite{lim2014}.}
\label{fig:geneva}
\end{figure}

\noindent In this Appendix, we present the calculations to derive Eq.~(\ref{main_eq}). To simplify the discussion, we have labeled different modes involved in the calculations in~Fig.~\ref{fig:geneva}. Suppose that the input state in mode $a$ is a coherent state $\ket{\sqrt{2\mu}}_a$ with creation operator $a^\dagger=(a_{\rm H}^\dagger+e^{i\phi_{\rm E}}a_{\rm V}^\dagger)/\sqrt{2}$. Also, suppose that the input signal in mode $b$ is the vacuum state $\ket{0}_b$. Then, the output signal in modes $c$ and $d$, after the action of the 50\,:\,50 beamsplitter (BS), is given by $\ket{\sqrt{\mu}}_c\otimes\ket{\sqrt{\mu}}_d$, where $c^\dagger=(c_{\rm H}^\dagger+e^{i\phi_{\rm E}}c_{\rm V}^\dagger)/\sqrt{2}$ and $d^\dagger=(d_{\rm H}^\dagger+e^{i\phi_{\rm E}}d_{\rm V}^\dagger)/\sqrt{2}$ denote the corresponding creation operators for modes $c$ and $d$. 

Next, we consider the phase modulator (PM) and the half-wave plate (HWP) that act on modes $c$ and $d$. The former performs the unitary transformation $c^\dagger=e^{i\varphi_{\rm B}}e^\dagger$, where $e^\dagger$ is the creation operator at the output port of the PM. The HWP applies the unitary transformation $d^\dagger_{\rm H}=f^\dagger_{\rm V}$ and $d^\dagger_{\rm V}=f^\dagger_{\rm H}$, where $f^\dagger_{\rm H}$ and $f^\dagger_{\rm V}$ denote the creation operators at the output port of the HWP. This means, in particular, that the quantum state in modes $e$ and $f$ has the form
\begin{equation}
\ket{\sqrt{\mu}{}e^{i\varphi_{\rm B}}}_e\otimes\ket{\sqrt{\mu}}_f,
\end{equation}
with the creation operators $e^\dagger$ and $f^\dagger$ given by $e^\dagger=(e_{\rm H}^\dagger+e^{i\phi_{\rm E}}e_{\rm V}^\dagger)/\sqrt{2}$ and $f^\dagger=(e^{i\phi_{\rm E}}f_{\rm H}^\dagger+f_{\rm V}^\dagger)/\sqrt{2}$, respectively.

Then, after applying the 50\,:\,50 BS on modes $e$ and $f$, we have that the output state in modes $g$ and $k$ can be expressed as
\begin{eqnarray}
\rm{exp}\bigg\{&&\frac{\sqrt{\mu}}{2}\Big[\big(e^{i\phi_{\rm E}}-e^{i\varphi_{\rm B}}\big)g^\dagger_{\rm H}+\big(1-e^{i(\phi_{\rm E}+\varphi_{\rm B})}\big)g^\dagger_{\rm V} \nonumber \\
&&+\big(e^{i\phi_{\rm E}}+e^{i\varphi_{\rm B}}\big)k^\dagger_{\rm H}+\big(1+e^{i(\phi_{\rm E}+\varphi_{\rm B})}\big)k^\dagger_{\rm V}\Big]\bigg\}\ket{0}. \nonumber \\
\end{eqnarray}
Finally, if we apply the polarising beamsplitters (PBS) (which we assume reflect horizontally polarised light and let vertically polarised light pass) on modes $g$ and $k$, we find that the state $\ket{\psi}$ at the input ports of Bob's detectors $\rm{D}_i$, with $i\in\{1,2,3,4\}$, is a tensor product of coherent states given by Eq.~(\ref{main_eq}).

\section{Side-channel attacks against Bob's linear optics network}\label{sca_linearopticsnetwork}

\noindent One main assumption of ddiQKD is that Bob's linear optics network [{\it i.e.,}\ the grey area within Bob's receiver in~Fig.~\ref{mdiQKD_vs_ddiQKD}(b)] is fully characterised and trusted. Note, however, that this does not mean that its devices need to be perfect, as this would be impossible to achieve in practice. In this Appendix we show that Eve could also exploit various typical imperfections of Bob's linear optics to avoid double clicks when performing the blinding attack described in the main text. 

\begin{figure}
\centerline{\includegraphics[width=0.95\columnwidth]{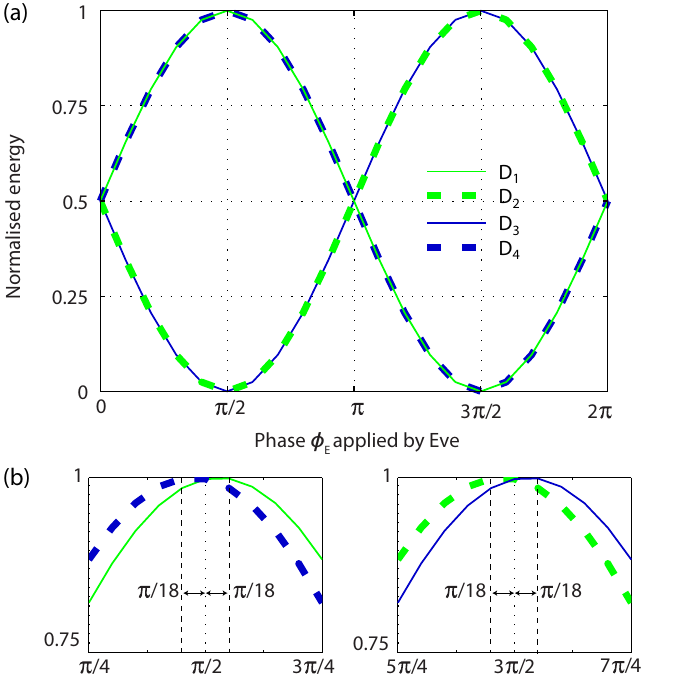}}
\caption{Normalised energy at the input ports of Bob's detectors ${\rm D}_i$ as a function of $\phi_{\rm E}$, when ${\bar \varphi}_{\rm B}=\pi/2$. (a)~Ideal scenario with a perfect PM that has $\Delta_{\varphi_{\rm B}}=0$. (b)~Example of a practical case where $\Delta_{\varphi_{\rm B}} = \pi/36$ \cite{xu2015}. The normalised energy is defined as the energy divided by the energy of a coherent state with mean photon number $\mu$. See text for further details.}
\label{fig:source}
\end{figure}

In particular, we will analyse two possible eavesdropping strategies in this context. In the first one, Eve uses the fact that Bob's PM $\varphi_{\rm B}$ is usually not perfect. More precisely, we study the situation where Bob's PM actually applies a phase $\varphi_{\rm B}={\bar \varphi}_{\rm B}+\Delta_{\varphi_{\rm B}}$, where ${\bar \varphi}_{\rm B}\in\{0,\pi/2,\pi,3\pi/2\}$ and the parameter $\Delta_{\varphi_{\rm B}}$ characterises the imperfection. In this scenario, Eve can select her phase $\phi_{\rm E}={\bar \phi}_{\rm E}+\Delta_{\phi_{\rm E}}$, where ${\bar \phi}_{\rm E}\in\{0,\pi/2,\pi,3\pi/2\}$ and $\Delta_{\phi_{\rm E}} > 0$ is a deviation term that Eve can select to control the detectors. According to Eq.~(\ref{main_eq}), the energy at the input ports of Bob's detectors ${\rm D}_1$, ${\rm D}_2$, ${\rm D}_3$ and ${\rm D}_4$ is proportional to, respectively, $\frac{\mu}{2}[1+\cos{(\phi_{\rm E}-\varphi_{\rm B})}]$, $\frac{\mu}{2}[1+\cos{(\phi_{\rm E}+\varphi_{\rm B})}]$, $\frac{\mu}{2}[1-\cos{(\phi_{\rm E}-\varphi_{\rm B})}]$, and $\frac{\mu}{2}[1-\cos{(\phi_{\rm E}+\varphi_{\rm B})}]$. For simplicity, below we focus on the case ${\bar \varphi}_{\rm B}=\pi/2$. The other cases can be analysed similarly. We consider first the ideal scenario where $\Delta_{\varphi_{\rm B}}=0$. The resulting normalised energies are illustrated in~Fig.~\ref{fig:source}(a) as a function of $\phi_{\rm E}$. That is, as already seen in the main text, when Eve's basis matches that of Bob, then two detectors receive maximum energy and, therefore, both click. If Bob and Eve use different bases then the total energy is equally distributed to all the four detectors and, given that $E_{\rm T}$ is chosen carefully, none of them click. Suppose now the practical scenario where Bob's state preparation is imperfect and $\Delta_{\varphi_{\rm B}}$ is equal to say, for instance, $\pi/36$ (or 5\textdegree, which is a typical accuracy in practical systems~\cite{xu2015}). In this situation, the energy distributions shift with respect to each other as highlighted in~Fig.~\ref{fig:source}(b). If ${\bar \phi}_{\rm E}=\pi/2$ and Eve selects say $\Delta_{\phi_{\rm E}}=\pi/18$ ($\Delta_{\phi_{\rm E}}=-\pi/18$) then the energy at the input ports of detectors ${\rm D}_1$ and ${\rm D}_4$ is, respectively, $E_+ \propto 0.998\mu$ and $E_- \propto 0.982\mu$ ($E_-$ and $E_+$). Similarly, if ${\bar \phi}_{\rm E}=3\pi/2$ the energy at the detectors ${\rm D}_2$ and ${\rm D}_3$ is, respectively, $E_-$ and $E_+$ ($E_+$ and $E_-$). That is, if Eve chooses carefully a suitable value of $\Delta_{\phi_{\rm E}}$ and $\mu$ such that $0.998\mu \geq \mu_{\rm th}$ and $0.982\mu < \mu_{\rm th}$, she can guarantee that only one detector clicks each given time, and no double-click is produced.

Finally, in the second eavesdropping strategy that we analyse we consider the situation where $\Delta_{\varphi_{\rm B}}=0$, but Eve exploits the fact that Bob's BSes are not perfect to avoid double-clicks. Although a 50\,:\,50 BS designed to operate at a certain wavelength (say, for example, at $1550$~nm) can achieve nearly perfect splitting ratio at that wavelength, its splitting ratio can vary significantly at a different wavelength. For instance, a custom-made beamsplitter sample studied in Ref.~\cite{li2011a} exhibited an extreme behaviour with splitting ratio of 98.6\,:\,1.4 (0.3\,:\,99.7) at $1470$~nm ($1290$~nm). While commercial beamsplitter models may exhibit less variation, Eve in general can to some extent control the splitting ratio by simply changing the wavelength of the signals~\cite{li2011a}, and this could be used to avoid double-clicks. 

In particular, suppose that Eve's signals are in a wavelength such that the splitting ratio of Bob's first (second) BS is $t_1:1-t_1$ ($t_2:1-t_2$). In addition, suppose that the creation operator of Eve's coherent states $\ket{\sqrt{2\mu}}$ is now given by $a^\dagger=(\sqrt{\gamma}{}a_{\rm H}^\dagger+e^{i\phi_{\rm E}}\sqrt{1-\gamma}a_{\rm V}^\dagger)$, where the parameter $\gamma$ is chosen by Eve. In this scenario, it can be shown that the state at the input ports of Bob's detectors $\rm{D}_i$ is a coherent state of the form 
\begin{eqnarray}\label{main_eq2}
\ket{\psi}&=&\ket{\alpha\big(\sqrt{{\hat t}_1{\hat t}_2{\hat \gamma}}e^{i\phi_{\rm E}}+\sqrt{t_1t_2\gamma}{}e^{i\varphi_{\rm B}}\big)}_{\rm{D}_1}\nonumber \\
&\otimes&\ket{\alpha\big(\sqrt{{\hat t}_1{\hat t}_2\gamma}+\sqrt{t_1t_2{\hat \gamma}}e^{i(\phi_{\rm E}+\varphi_{\rm B})}\big)}_{\rm{D}_2} \nonumber \\
&\otimes&\ket{\alpha\big(\sqrt{{\hat t}_1t_2{\hat \gamma}}e^{i\phi_{\rm E}}-\sqrt{t_1{\hat t}_2\gamma}{}e^{i\varphi_{\rm B}}\big)}_{\rm{D}_3} \nonumber \\
&\otimes&\ket{\alpha\big(\sqrt{{\hat t}_1t_2\gamma}-\sqrt{t_1{\hat t}_2{\hat \gamma}}e^{i(\phi_{\rm E}+\varphi_{\rm B})}\big)}_{\rm{D}_4}, 
\end{eqnarray} 
where ${\hat x}=1-x$, and $\alpha=\sqrt{2\mu}$. Note that when $t_1=t_2=\gamma=1/2$ we obtain Eq.~(\ref{main_eq}). 

\begin{figure}
  \includegraphics[width=0.9\columnwidth]{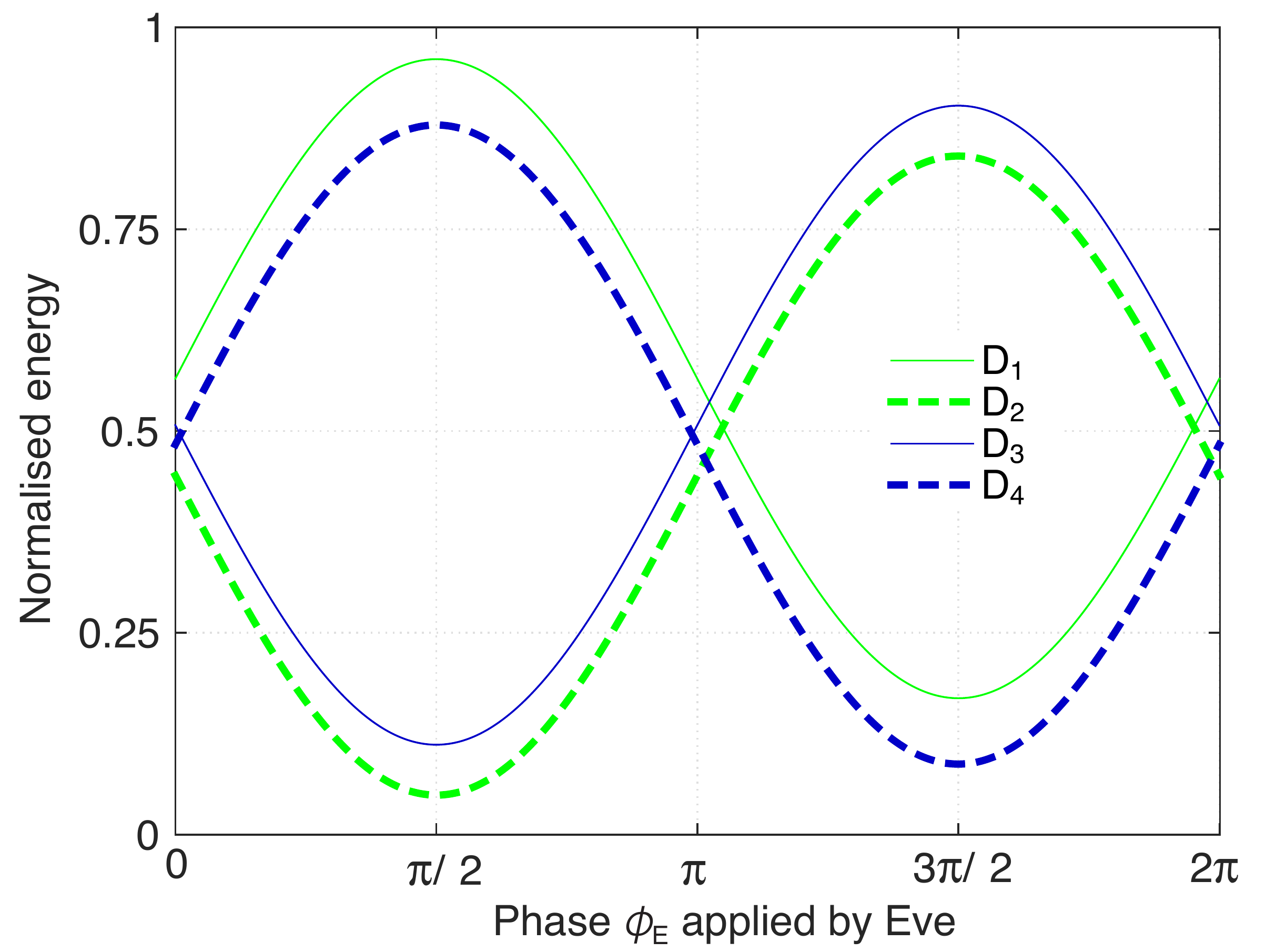}
  \caption{Normalised energy at the input ports of Bob's detectors ${\rm D}_i$ as a function of $\phi_{\rm E}$, when $\varphi_{\rm B}=\pi/2$. Here we assume that the splitting ratio of Bob's first (second) BS is 44\,:\,56 (46\,:\,54), and Eve's state parameter $\gamma=0.2$. See text for further details.}
  \label{fig:int_mismatch_phpi_2}
\end{figure}

This means that, in principle, Eve might select the parameter $\gamma$ and the wavelength of her signals such that the resulting splitting ratios $t_1$ and $t_2$ make the input energies at Bob's detectors asymmetric. In so doing, and following a similar argumentation to the one introduced in the previous eavesdropping strategy, Eve can guarantee that when she and Bob choose the same basis, only one detector clicks. This situation is illustrated in~Fig.~\ref{fig:int_mismatch_phpi_2} for a particular example where $\varphi_{\rm B} = \pi/2$, $t_1=0.44$, $t_2=0.46$, and $\gamma=0.2$. In this scenario, we find that the maximum normalised energy at the input ports of Bob's detectors ${\rm D}_1$ and ${\rm D}_4$ when Eve selects $\phi_{\rm E}=\pi/2$ is, respectively, $0.96$ and $0.87$. Similarly, when she chooses $\phi_{\rm E}=3\pi/2$ the maximum normalised energy at the detectors ${\rm D}_3$ and ${\rm D}_2$ is, respectively, $0.9$ and $0.84$. Therefore, Eve can choose the energy of her signals such that only the detector ${\rm D}_1$ (${\rm D}_3$) clicks when $\phi_{\rm E}=\pi/2$ ($\phi_{\rm E}=3\pi/2$). That is, by changing the values of the parameters $t_1$, $t_2$, and $\gamma$, Eve can guarantee that only one detector clicks each given time. 

\def\bibsection{\medskip\begin{center}\rule{0.5\columnwidth}{.8pt}\end{center}\medskip} 

%

\end{document}